\documentclass[conference]{IEEEtran}
\IEEEoverridecommandlockouts
\usepackage{cite}
\usepackage{amsmath}
\usepackage{flexisym}
\usepackage{breqn}
\usepackage{amssymb,amsfonts}
\usepackage{caption}
\usepackage{supertabular,booktabs}
\usepackage{algorithmic}
\usepackage{graphicx}
\usepackage{booktabs}
\usepackage{subcaption}
\usepackage{textcomp}
\usepackage{longtable}
\usepackage{xcolor}
\def\BibTeX{{\rm B\kern-.05em{\sc i\kern-.025em b}\kern-.08em
    T\kern-.1667em\lower.7ex\hbox{E}\kern-.125emX}}
\begin{document}

\title{A Framework to Evaluate Independent Component Analysis applied to EEG signal: testing on the Picard algorithm
\thanks{Funding provided by The Swartz Foundation (Oldfield, NY) and by the  National Institute of Neurological Disorders and Stroke, U.S.A. (R01-NS047293) }
}

\author{\IEEEauthorblockN{Gwenevere Frank}
\IEEEauthorblockA{\textit{Electrical and Computer Engineering} \\
\textit{University of California San Diego}\\
La Jolla, USA \\
jfrank@ucsd.edu}
\and
\IEEEauthorblockN{Scott Makeig}
\IEEEauthorblockA{\textit{Institute for Neural Computation} \\
\textit{University of California San Diego}\\
La Jolla, USA \\
smakeig@ucsd.edu}
\and
\IEEEauthorblockN{Arnaud Delorme}
\IEEEauthorblockA{\textit{Swartz Center for Computational Neuroscience,} \\\textit{Institute for Neural Computation} \\
\textit{University of California San Diego}\\
La Jolla, USA \\
\textit{Centre de recherche Cerveau et Cognition} \\
\textit{Paul Sabatier University}\\
Toulouse, France \\
arnodelorme@gmail.com}
}

\maketitle

\begin{abstract}
Independent component analysis (ICA), is a blind source separation method that is becoming increasingly used to separate brain and non-brain related activities in electroencephalographic (EEG) and other electrophysiological recordings. It can be used to extract effective brain source activities and estimate their cortical source areas, and is commonly used in machine learning applications to classify EEG artifacts. Previously, we compared results of decomposing 13 71-channel scalp EEG datasets using 22 ICA and other blind source separation (BSS) algorithms. We are now making this framework available to the scientific community and, in the process of its release are testing a recent ICA algorithm (Picard) not included in the previous assay. Our test framework uses three main metrics to assess BSS performance: Pairwise Mutual Information (PMI) between scalp channel pairs; PMI remaining between component pairs after decomposition; and, the complete (not pairwise) Mutual Information Reduction (MIR) produced by each algorithm. We also measure the “dipolarity” of the scalp projection maps for the decomposed component, defined by the number of components whose scalp projection maps nearly match the projection of a single equivalent dipole located in the volume of a template boundary element method (BEM) electrical forward problem head model. Within this framework, Picard performed similarly to Infomax ICA. This is not surprising since Picard is a type of Infomax algorithm that uses the L-BFGS method for faster convergence, in contrast to Infomax and Extended Infomax (runica) which use gradient descent. Our results show that Picard performs similarly to Infomax and, likewise, better than other BSS algorithms, excepting the more computationally complex AMICA. Further research might determine if partial Picard decomposition, followed by AMICA, might produce unequaled performance without a large time penalty. We have released the source code of our framework and the test data through GitHub to encourage further comparisons of ICA/BSS algorithm performance applied to electrophysiological data (https://github.com/sccn/testica).
\end{abstract}


\section{Introduction}
Scalp-recorded EEG data generated from brain signals is generally thought to be the result of synchronous local field potential (LFP) activity arising from coherent or near-coherent local field activity within patches of adjacent, radially oriented cortical pyramidal cells \cite{nunez1974brain, varela2001brainweb}. Multiple facets of biology support the fact that under suitable conditions ICA should be capable of separating activities resulting from far-field projection to the scalp of cortical local field activity coherent within compact, physically distinct cortical areas producing temporally near independent contributions across scalp recording channels:
First, connections between neurons in close proximity ($<<100 \mu m$) are far denser than those between neurons located further from each other \cite{stepanyants2009fractions, stettler2002lateral}, and inhibitory and glial cell networks have no long range connectivity \cite{stepanyants2009fractions}.
Second, connections between the thalamus and cortex that play a significant role in cortical dynamics are primarily radial connections \cite{sarnthein2005thalamocortical, dehghani2010magnetoencephalography}.

For the above reasons, the majority of the LFPs contributing to EEG signals recorded at the scalp should be generated by coherent field activity emerging within small dense cortical “patches” in which short-range connections dominate, rather than in more widely dispersed, at best tenuously connected networks. As a result, effective sources of EEG signals recorded at the scalp should be primarily the mixture, at the scalp electrodes, of synchronous or near synchronous activity in such compact locally connected cortical “patches”, volume conducted and mixed linearly  at the scalp electrodes.

\section{Background}

\subsection{Algorithm comparison framework}
We used 3 metrics to compare algorithms:

\subsubsection{Pairwise Mutual Information (PMI)}

Pairwise Mutual Information measures the mutual information between (any or all possible) pairs of row or column vectors in two matrices. That is, PMI is defined by 
\begin{dmath}
[M]_{ij} = I(x_{i}; x{j})=h(x_{i})+h(x_{j})-h(x_{i},x_{j}),  i,j \in 1, \dots, n
\end{dmath}
Where $x$ is a vector of length $N$ and each element in $x$ is a time series $x_{i}(t)$ of length $n$ and $M$ is a $nxn$ matrix.
For the purposes of this study, entropy values were computed by constructing a histogram and estimating the integral via a Riemann approximation. This technique is generally considered to yield good approximations when sample sizes are large, which is in fact the case for EEG data. It is also possible to approximate the asymptotic variance of the estimated entropy values using information from the constructed histogram, allowing for the analysis of the statistical significance of the results.

Let $B$, the number of bins in the histogram, be some fixed value of our choosing. We then will denote the histogram of $x_{i}$ as $b_{i}(k), k \in 1,..B$, specifically $b_{i}(k)$ is the number of samples for which the value $x_{i}(t)$ is in the $kth$ bin. $p(x_{i})$ can then be estimated as $b_i(k)/(N \Delta_{k})$ where $\Delta_k$ is the bin width of bin $k$ and $N$ is the total number of bins. Therefore, the Riemann approximation of the integral of the continuous density function correctly integrates to one.
\begin{dmath}
$$\sum_{k} p(x) \Delta_{k} = \sum_{k} (b_{i}(k)/(N \Delta_{k})) \Delta_{k} = 1$$
\end{dmath}
We may then write the estimate of marginal entropy as
\begin{dmath}
$$\sum_{k} -p(x)log(p(x)) \Delta_{k} = - \sum_{k} (b_{i}(k)/(N \Delta_{k})) log(b_{i}(k)/(N \Delta_{k})) \Delta_k = H_i - log(B)$$
\end{dmath}
Where $H_i$ is the entropy of the discrete probability distribution given by $b_{i}(k)/(N), k \in 1, \dots, B$. We then estimate the joint entropy in a similar manner. $p(x_{i},x_{j})$ is estimated by $b_{ij}(k,l)/(N \Delta_{k} \Delta_{l})$ where $b_{ij}(k,l)$ is the number of time points $t$ for which $x_{i}(t)$ is in bin $k$ and $x_{j}(t)$ is in bin $l$. We then may write the estimate of the joint entropy as 
\begin{dmath}
\sum_{k} \sum_{l} - p(x_{i},x_{j}) log( p(x_{i}, x_{j}) ) \Delta_{k} \Delta{l} = - \sum_{k} \sum_{l} (b_{ij}(k,l)/N) log (b_{ij}(k,l)/N) - 2log(B)
\end{dmath}
We further define
\begin{dmath}
H_{ij} = - \sum_{k} \sum_{l} (b_{ij}(k,l)/N) log (b_{ij}(k,l)/N) 
\end{dmath}
Finally, we may define $M_{ij}$, the mutual information between $x_{i}$ and $x_{j}$, as 
\begin{dmath}
$$M_{ij} = (H_{i} - log(B)) + (H_{j} - log(B)) - (H_{ij} - 2 log (B)) = H_{i}+H_{j}-H{ij}$$
\end{dmath}
Of particular note, the resulting estimate of mutual information does not depend on bin size, including not depending on the fact that equal bin sizes were used. For the PMI calculations in this study the diagonal entries of $M$ were set to zero.

\subsubsection{Mutual Information Reduction (MIR)}

The reduction in mutual information that results from applying an unmixing matrix $W$ to the data $x$ can be computed relatively easily using only one-dimensional density models, as pointed out by J. Palmer in \cite{delorme2012independent}. MIR, or Mutual Information Reduction, can be defined as follows
\begin{dmath}
$$ MIR = I(x) - I(y) = [h(x_1) + \dots + h(x_n)] - [h(y_1) + \dots + h(y_n)] - h(x) + log |det W| + h(x) = log |det W| + [h(x_1) + \dots + h(x_n)] - [h(y_1) + \dots + h(y_n) ] $$
\end{dmath}
Intuitively, this value represents how much mutual information is removed from the data by performing ICA/BSS. The above formulation for MIR depends only on the log of the determinant of $W$ and the marginal entropies of $x$ and $y$. The marginal entropy values can be estimated using histograms in a similar procedure as that outlined for computing PMI. ICA seeks to minimize mutual independence and thereby maximize MIR. We can expect that the ICA algorithms that produce the most independent sources, and therefore minimize remnant mutual information, will give the highest MIR values, although the produced MIR values can vary widely across datasets.

\subsubsection{Equivalent dipole modeling}
For each algorithm, a best fitting single equivalent dipole model was found for each decomposed independent component (IC). Dipoles were found using EEGLAB function DIPFIT using a spherical four-shell BEM head model (radius: 71, 72, 79, 85 mm; shell conductances: 0.33, 0.0042, 1, 0.33 mS). Due to the simplicity of this template head model, to minimize errors the channels including the two electrodes nearest the eyes were excluded from dipole fitting. It is worth noting that fitting a single dipole to model each component is somewhat simplistic, as some ICA components clearly better fit a model of synchronous activity from two bilaterally near-symmetric sources. However, components that clearly would be better modeled by a two dipole model appeared rare in this data. In particular, components accounting for EOG-related activity in the data would best be modeled using a two dipole solution. However, given the difficulty in constructing an accurate forward model of the front of the skull, the relatively small distance between the eyes makes it unlikely that error caused by modeling these components using a single (midline) dipole would be large.

\subsection{Algorithm selection}

\subsubsection{Original algorithms}
Performance of 20 blind source separation algorithms were compared, including PCA (principal component analysis) and 19 ICA/BSS algorithms. MATLAB implementations of these algorithms are publicly available from various sources, as listed in the origins column of Table 1.

\subsubsection{The Picard algorithm}
Picard \cite{ablin2018faster} is a relatively new ICA algorithm that has already been applied in neuroscience studies \cite{chaisaen2020decoding, sangnark2021revealing, kudo2022magnetoencephalography}. Picard’s novelty is in its use of the L-BFGS algorithm for minimization of its objective function, and in preconditioning with an approximation of the hessian matrix.
Picard solves the ICA problem in the maximum likelihood sense; that is, Picard seeks to find $W$ such that $W$ minimizes the negative log likelihood
\begin{dmath}
\mathcal{L}(W) = -log |det(W)| - \hat{E} [ \sum_{i=1}^{N}log(p_{i}(y_{i}(t))]
\end{dmath}
Where $\hat{E}$ is the sample average.
To minimize the loss function, Picard employs an optimization strategy called L-BFGS \cite{liu1989limited}.
L-BFGS is a memory optimized approximation of the popular BFGS optimization algorithm that belongs to the family of quasi-newton methods that estimate the curvature of a loss function without computing the hessian. L-BFGS starts with an initial guess of the hessian that is iteratively refined at each step of optimization by incorporating updates from the last $m$ steps ($m$ is called the memory parameter). The initial guess of the hessian must be easily invertible; standard L-BFGS uses a multiple of the identity matrix. However, Picard uses an approximation of the hessian (see \cite{ablin2018faster} for details on the computation of this approximation) as a more accurate starting guess, effectively preconditioning the problem.
L-BFGS also requires a search at each iteration to determine an appropriate step size $\alpha$. To reduce computation cost, Picard uses a relatively simplistic backtracking strategy for this.

Picard's loss function may also be tweaked such that Picard solves the same optimization problem as Extended Infomax, which explicitly looks for sources with sub-kurtotic and well as super-kurtotic probability distributions. Here we used Picard version mimicking original (non-Extended) ICA. There also exists an orthogonal variant of Picard, deemed Picard-O \cite{ablin2018ortho}, that solves the same optimization problem as the FastICA algorithm.

\subsection{EEG Data}
Fourteen participants were selected from a pool of volunteers and asked to perform a visual working memory task \cite{onton2005frontal}. Each trial begins with a 5-sec fixation period during which a fixation cross was displayed at the center of the screen. Next, participants were presented at near-sec intervals with a series of single letters, some (black) to be memorized and some (green) to be ignored. Participants were instructed to then retain the identities of the to-be-memorized letters for a few seconds. They were then presented a single probe letter and had to indicate whether or note it was in the memorized letter set by pressing the appropriate button. 400 ms following the button press, participants received auditory feedback informing them if they were correct or not. Each subject participated in 100-150 trials. Additional information on the experimental paradigm and associated event-related analysis can be found in Onton et al. \cite{onton2005frontal}. Data from human participants was collected using an experimental protocol approved by the Institutional Review Board of University of California San Diego. Each participant gave written consented to participate.

A 71-channel EEG system was used for recording.  All channels were referenced to the right mastoid.  Data were recorded at 250 Hz and subsequently band-pass filtered using an analog filter with a pass band of 0.01 to 100 Hz. The participant's scalp was prepared to enable all channels  to have an input impedance below 5 k$\Omega$. 13 participants' data were selected to be included in an analysis based on visual inspection of the quality of its ICA decomposition. Additional information on the criteria used for the inclusion and exclusion of participants can be found in \cite{delorme2012independent}.
Data processing was performed using a custom processing pipeline written in MATLAB using the EEGLAB environment \cite{delorme2004eeglab}. Recorded data were first high pass filtered using a FIR filter with a 0.5-Hz cutoff. Then the data were epoched, each extracted epoch from 700 ms before to 700 ms after a letter presentation onset. Mean channel values were subtracted from each epoch, and noisy epochs were rejected by visual inspection. Abnormal high amplitude and high frequency data content (such as the activity indicative of jaw clenching or sneezing) was the primary criteria used for rejection of noisy epochs. Between 1 and 16 epochs were rejected per participant. After this initial processing, the epoched dataset for each subject included 269,000 and 315,000 data points.

\section{Results}

\subsection{Algorithm Comparison}

\begin{table}[ht]
\setlength{\LTpost}{0mm}
\caption{{\large MIR Results}\label{tableRes}}

\begin{tabular}{lrrl}
\toprule
\textbf{Algorithm} & \textbf{MIR (Kbits/s)} & \textbf{ND 5\%} & Source \\ 
\midrule
Amica & $43.13$ & $30.01$ & EEGLAB 6.1 \\ 
Infomax & $43.07$ & $25.35$ & EEGLAB 4.515 \\ 
\color{red} \emph{Picard} & $43.06$ & $25.14$ & Github* \\ 
Ext. Infomax & $43.02$ & $25.24$ & EEGLAB 4.515 \\ 
Pearson & $43.01$ & $25.89$ & ICAcentral (6) \\ 
\color{red} \emph{Picard-O} & $43.00$ & $25.03$ & Github* \\ 
SHIBBS & $42.74$ & $18.96$ & ICAcentral (5) \\ 
JADE & $42.74$ & $18.42$ & EEGLAB 4.515 \\ 
FastICA & $42.71$ & $20.15$ & ICAcentral (2) \\ 
TICA & $42.68$ & $17.23$ & ICALAB 1.5.2 \\ 
JADE opt. & $42.64$ & $14.84$ & ICALAB 1.5.2 \\ 
SOBI & $42.51$ & $12.46$ & EEGLAB 4.515 \\ 
JADE-TD & $42.47$ & $13.43$ & ICALAB 1.5.2 \\ 
SOBIRO & $42.44$ & $13.43$ & EEGLAB 4.515 \\ 
FOBI & $42.31$ & $10.73$ & ICALAB 1.5.2 \\ 
EVD24 & $42.30$ & $10.40$ & ICALAB 1.5.2 \\ 
EVD & $42.19$ & $9.64$ & ICALAB 1.5.2 \\ 
icaMS & $42.18$ & $7.48$ & ICA DTU Tbox \\ 
AMUSE & $42.14$ & $5.74$ & ICALAB 1.5.2 \\ 
PCA & $41.86$ & $3.79$ & EEGLAB 4.515 \\ 
\bottomrule

\end{tabular}

\begin{minipage}{\linewidth}
Column one gives the name of the algorithm. Column two (Mutual Information Reduction, MIR, in kilobits per second) lists the mean reduction in mutual information achieved by the (component) decomposition compared to the original (scalp channel) data. The third column (near-dipolar percentage, ND 5\%) gives the percentage of produced components whose scalp projection maps have less than 5\% residual variance from the scalp projection pattern of the best fitting single equivalent model dipole. The fourth column (Origin) list the names of the software packages that provided the implementation used for each decomposition algorithm. The implementations of Picard and Picard-O used were the latest available from https://github.com/pierreablin/picard (commit 1557b10)\\
\end{minipage}

\end{table}

\begin{figure}[h]
\centering
\includegraphics[width=0.5\textwidth]{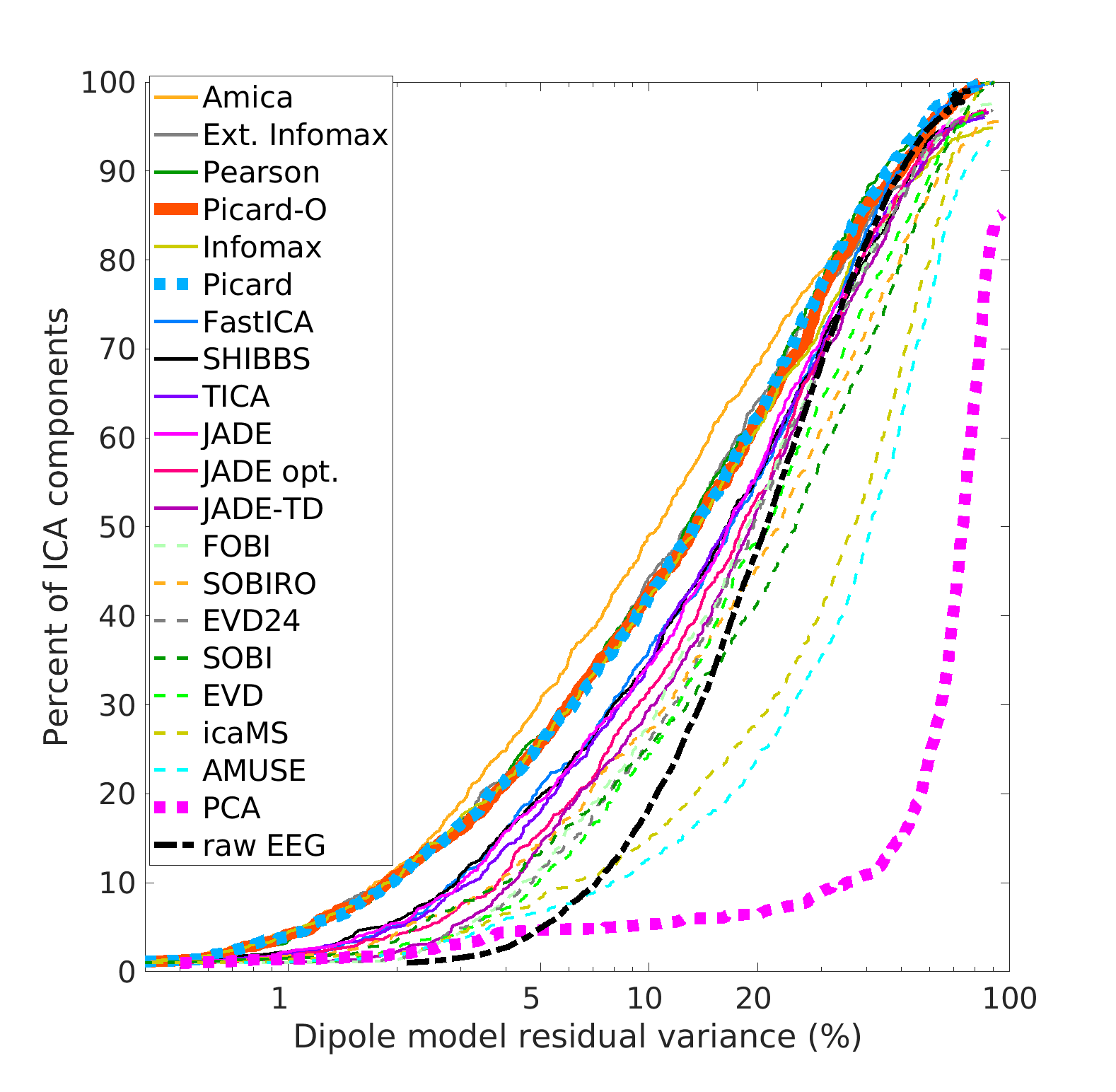}
\caption{Decompositions that reduce total mutual information in the component time courses also return more components with near-dipolar scalp maps.}
\label{rv_vs_percentICA}
\end{figure}

\begin{figure*}
\centering
\begin{subfigure}[b]{0.49\textwidth}
\centering
\includegraphics[width=\textwidth]{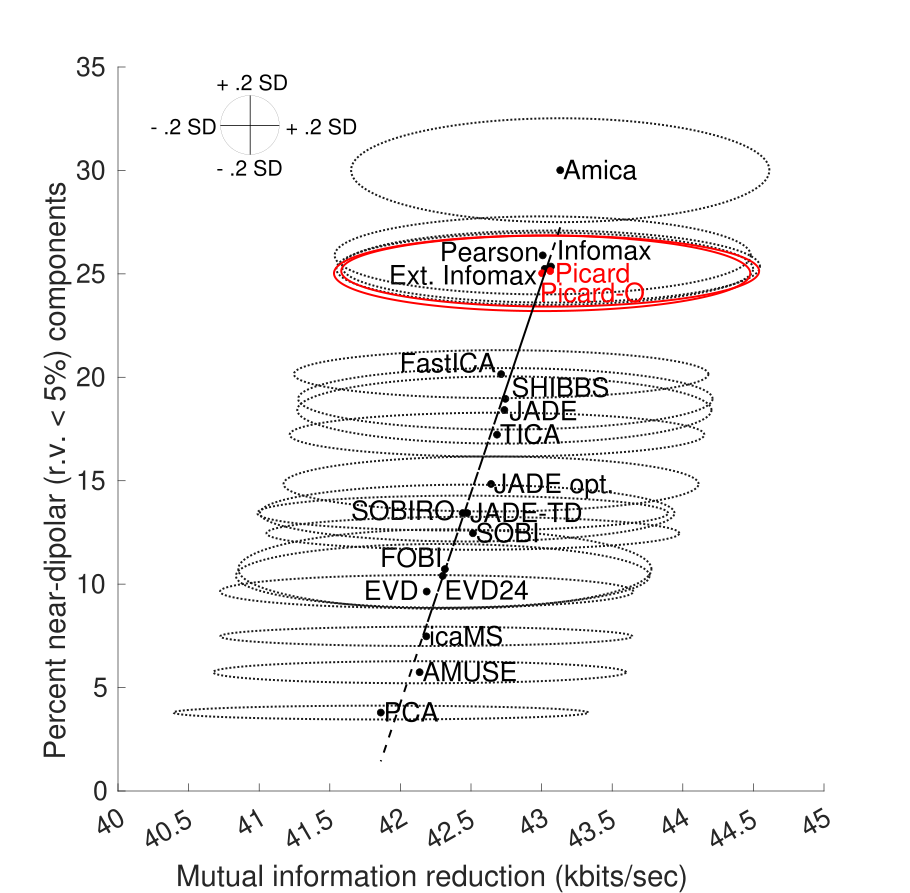} 
\caption{MIR vs Dipolarity} \label{fig:data}
\end{subfigure}
\hfill
\begin{subfigure}[b]{0.49\textwidth}
\centering
\includegraphics[width=\textwidth]{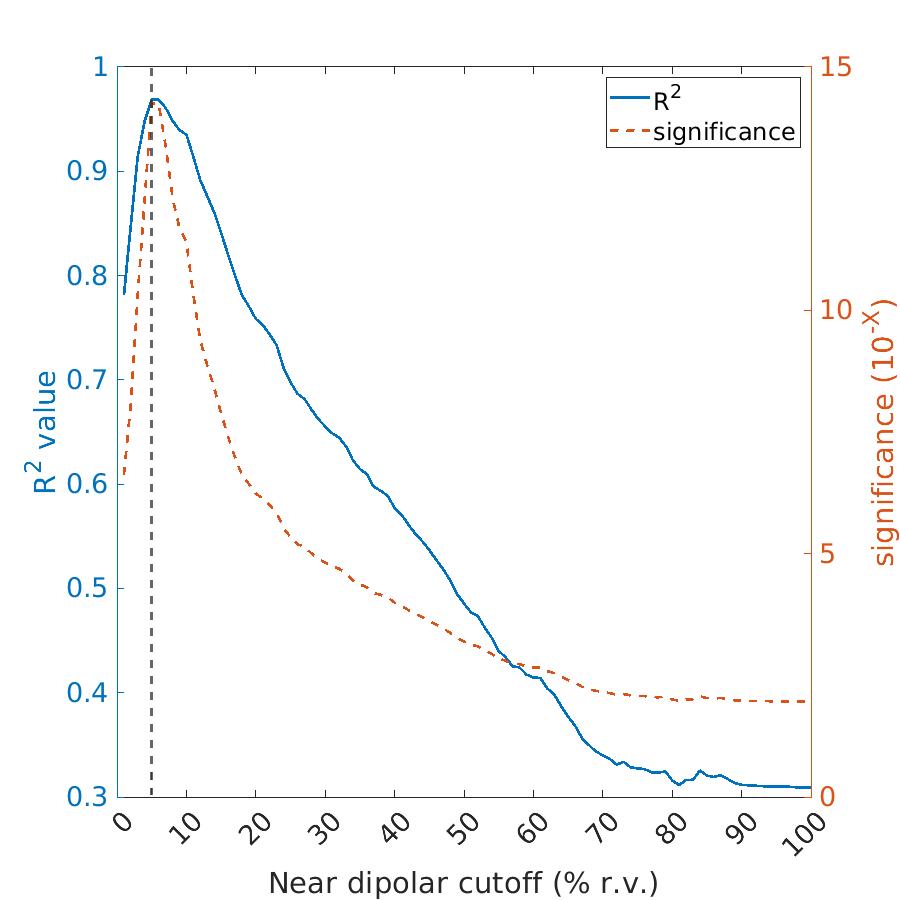} 
\caption{MIR $R^2$ and Significance} \label{fig:stats}
\end{subfigure}

\begin{subfigure}[b]{0.49\textwidth}
\centering
\includegraphics[width=\textwidth]{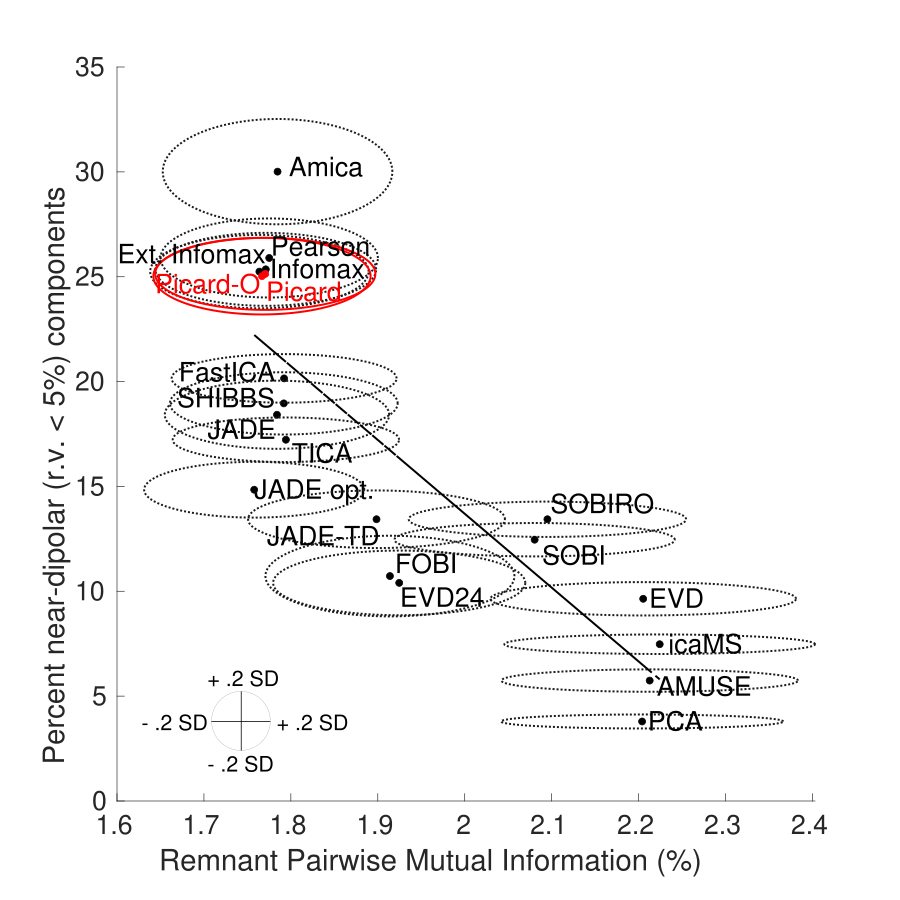} 
\caption{MIR vs Dipolarity} \label{fig2:PMIdata}
\end{subfigure}
\begin{subfigure}[b]{0.49\textwidth}
\centering
\includegraphics[width=\textwidth]{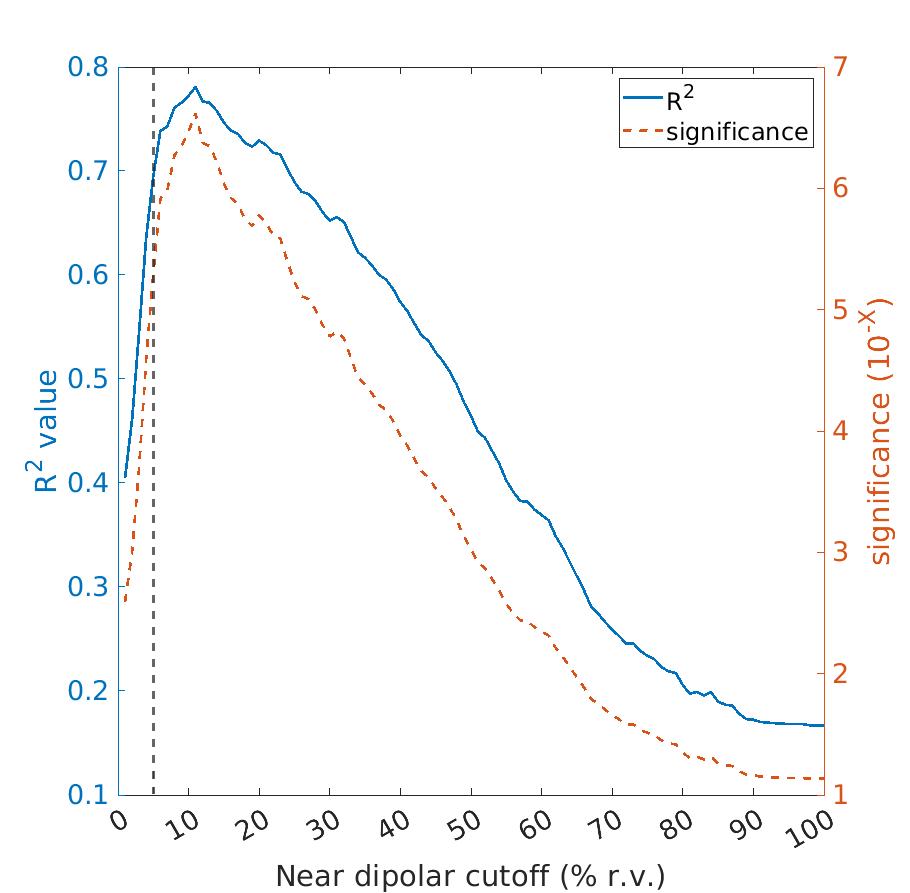} 
\caption{PMI $R^2$ and Significance} \label{fig2:PMIstats}
\end{subfigure}

 \caption{ Panel \subref{fig:data} shows percentage of near-dipolar components produced versus MIR for the selected algorithms. Picard as well as Picard-O and their associated errors are plotted in red. Ellipses around data points represent $\pm 0.2$ standard deviations for MIR and $\pm 0.2$ standard deviation for near-dipolar percentage. Panel \subref{fig:stats} shows (y axis) the $R^2$ (left) and significance ($\alpha$) (right) values  for linear regressions of MIR on percentage of near-dipolar components across (x axis) a range of 'near-dipolar' r.v. thresholds. The vertical dashed line indicates a 5\% r.v. cutoff. Significance is scaled by $-log_{10}$. Panel (\subref{fig2:PMIdata}) shows percentage near-dipolar components (y axis) versus remnant component Pairwise Mutual Information (x axis) for the 20 BSS algorithms. Picard as well as Picard-O and their associated data points are plotted in red. Ellipses around data points represent $\pm 0.2$ standard deviations for percent remnant PMI and for percentage of near-dipolar components. Panel \subref{fig2:PMIstats} shows (y axis) the $R^2$ (left) and significance ($\alpha$) values (right) for linear regressions of the percentage of remnant component PMI vs the percentage of produced near-dipolar components across  a range of 'near-dipolar' r.v. thresholds. The vertical dashed line indicates a 5\% r.v. cutoff. Significance is scaled by $-log_{10}$.}

\end{figure*}

\begin{figure}[h]
\centering
\includegraphics[width=0.5\textwidth]{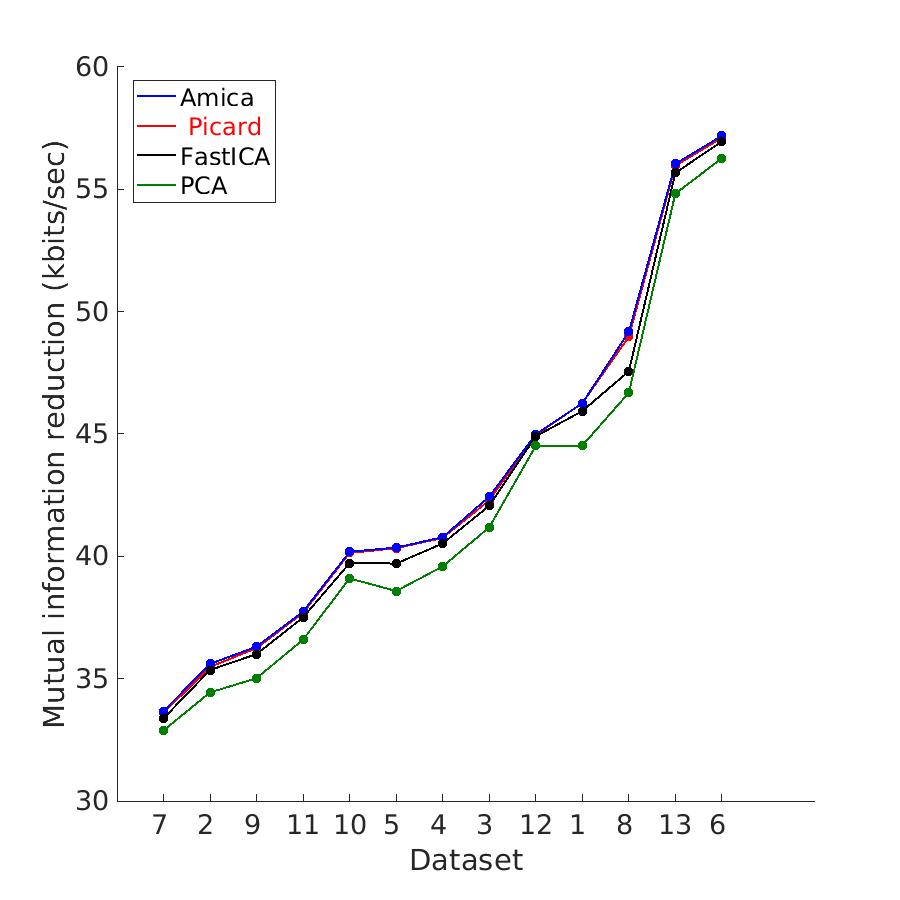}
\caption{Comparison of mean Mutual Information Reduction (MIR) produced by AMICA, Picard, FastICA, and PCA transformations of the data from channel activities to component activations for the 13 analyzed datasets, ordered by increasing mean MIR. Note that the order of the three algorithms is preserved across datasets even though the different datasets give quite different MIR values.}
\label{dataset_comp}
\end{figure}

\begin{figure}[h]
\centering
\includegraphics[width=0.5\textwidth]{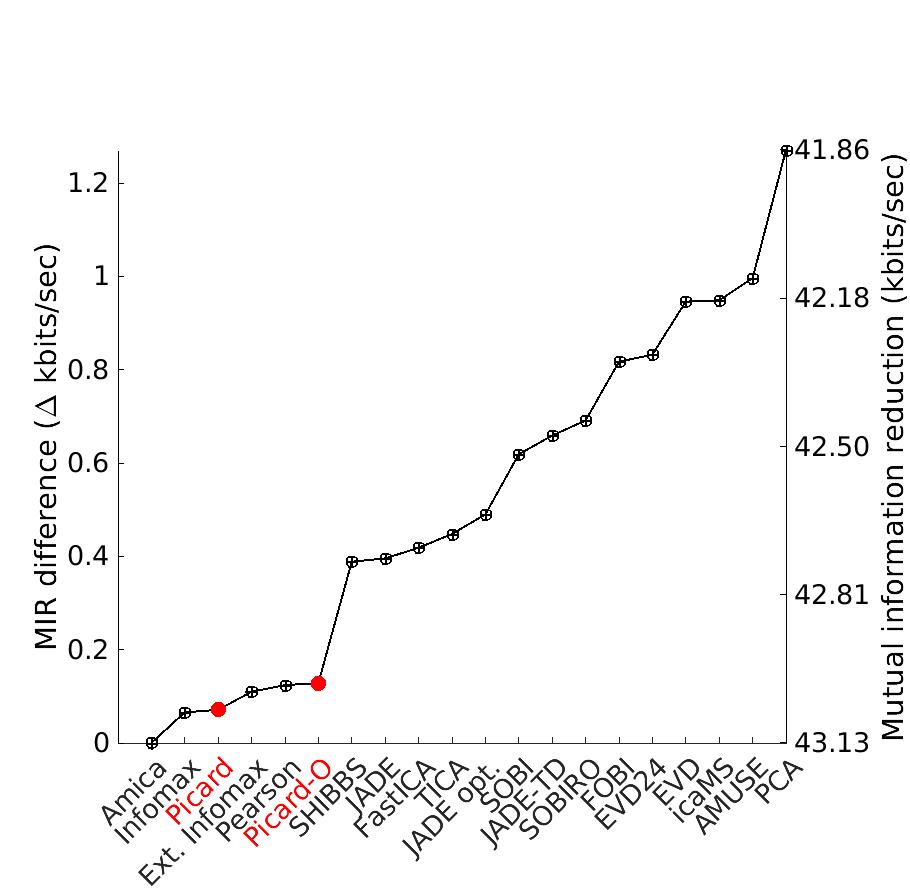}
\caption{MIR difference from AMICA (far left) for each algorithm tested. Picard and Picard-O are plotted in red.}
\label{MI_difference}
\end{figure}

\begin{figure}[h]
\centering
\includegraphics[width=0.5\textwidth]{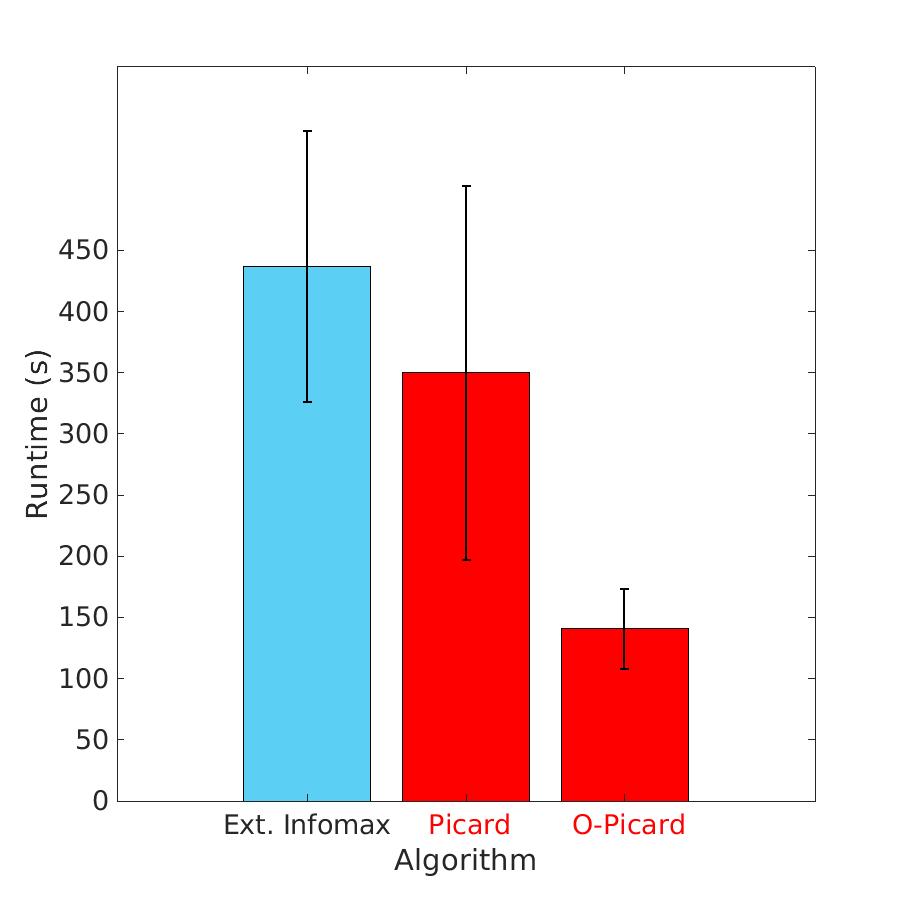}
\caption{Average run time across 5 decompositions of each dataset by Extended Infomax, Picard, and O-Picard.}
\label{runtime}
\end{figure}

Table \ref{tableRes} gives the mean MIR for each algorithm averaged across 13 datasets, as well as the percentage of near-dipolar components using a $5\%$ residual variance (r.v.) threshold.

Figure \ref{rv_vs_percentICA} shows (y axis) the mean percentage of components with residual scalp map variance (after subtracting the computed scalp projection map of the best-fitting single equivalent model dipole) less than the (x axis) indicated value.
The decomposition methods listed in the legend are sorted by mean number of near-dipolar components (using a $\leq 10 \%$ dipolarity threshold).
Of particular note, Picard is shown by a teal dashed line, Picard-O by a solid orange line, PCA by a pink dashed line, and the dipolarity distribution of the raw EEG data (for 71 randomly selected EEG time point scalp maps) by the black dashed line. As might be expected, the performance of Picard is quite similar to that of Infomax, while AMICA gives the best overall performance for dipolarity r.v. thresholds from about $\leq 3\%$ to $\leq 30\%$.

Figure \ref{fig:data}  compares the percentage of near-dipolar components (using an r.v. threshold of $\leq 5\%$), versus mean Mutual Information Reduction for each algorithm. A linear regression of the data points is also plotted as a dashed line; the fit to a straight line is remarkable ($R^{2}=0.97$, $\alpha=5.6*10^{-15}$). 

Figure \ref{fig:stats} shows both (left) the probability that the percentage of near-dipolar components varies linearly with MIR (plotted in blue) and the percentage of the total variance in the data that can be explained by the MIR versus percent near-dipolar relationship (y axis, plotted in red) versus (x axis) the percentage r.v. value used as the near-dipolar component threshold. Of particular note, both of these values peak at a r.v. cutoff value of approximately 6 percent. Again the performance of Picard is quite similar to that of Infomax.

Figure \ref{fig2:PMIdata} compares the percentage of returned near-dipolar components, defined as having a scalp map w.r.t. the computed scalp projection of the equivalent dipole model r.v. $\leq 5\%$, versus remaining component Pairwise Mutual Information (PMI) for each algorithm. Remnant PMI is computed as the percentage the mean component PMI represents of mean scalp channel PMI (means of all elements of the PMI matrices, averaged across datasets). A linear regression of these data (again, significant: $R^{2}=0.69$, $\alpha=5.06*10^{-6}$) is shown by a dashed line. 

Figure \ref{fig2:PMIstats} shows both (in blue) the probability that the percentage of near-dipolar components varies linearly with remnant PMI and (in red) the percentage of total data variance that can be explained by the linear  relationship versus (x axis) the near-dipolar r.v. threshold value. Again, the performance of Picard is close to that of Infomax.

Figure \ref{dataset_comp} shows the MIR for each of the datasets produced by AMICA (blue), Picard (red), FastICA, and PCA. The order of the algorithms is maintained for nearly every dataset, although the differences between datasets dwarf those between algorithms.  

Figure \ref{MI_difference} plots two trends together, MIR difference from AMICA (on the left) and mean MIR (on the right y axis). MIR difference from AMICA (plotted in blue) is computed as the superiority in mean MIR for AMICA relative to the other algorithms, which are sorted by mean Mutual Information Reduction.

Figure \ref{runtime} plots the mean run times of three of the algorithms: Extended Infomax, Picard, and Orthogonal Picard. Runtime measurements were conducted on the San Diego Supercomputer Center's Expanse system with an allocation of 16 CPU cores and 64 GB of memory. Each bar represents the average wall clock time for the respective algorithm to perform the decomposition, averaged across five iterations of each dataset. All algorithms were run using a maximum of 100,000 steps. Extended Infomax used a stopping rule step size threshold of $1*10^{-8}$, Picard and Picard-O a stopping rule step size of $1*10^{-6}$. 

\subsection{Picard Tuning}

\begin{figure}[h]
\centering
\includegraphics[width=0.5\textwidth]{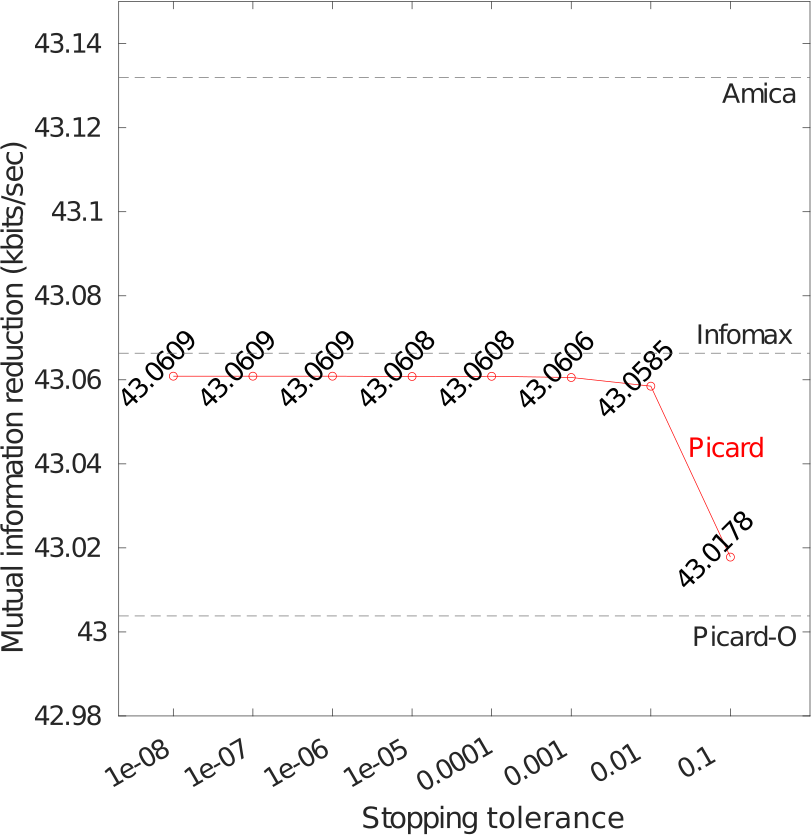}
\caption{Average Mutual Information Reduction (MIR) for Picard across a series of stopping rule step size thresholds. Mean MIR returned by AMICA, Infomax, and Picard-O are shown as dashed horizontal lines. Improvement for stopping rule criteria less than $10^{-3}$ is minimal.}
\label{mir_vs_stoppingRule}
\end{figure}

Figure \ref{mir_vs_stoppingRule} compares the MIR achieved by Picard across different stopping rule step sizes $(10^{-1},10^{-2},10^{-3},10^{-4},10^{-5},10^{-6},10^{-7},10^{-8})$. The plotted values are the mean MIR for each stopping rule threshold averaged across all datasets. Horizontal dashed lines show mean MIR (across datasets) for AMICA, Infomax, Pearson, and Picard-O. Stopping rule thresholds smaller than $10^{-3}$ produce very little further MIR improvement.



\section{Discussion}
\subsection{Physiological plausibility of ICA}
 EEG channels measure a time series of potential differences between two or more scalp sensors. For analysis, is common to transform the data to “average reference,” making each channel represent a time series of potential difference between a given electrode and the simultaneous mean potential across all electrodes. EEG scalp channels measure a linear mixture of the far-field projections of electrical sources both outside the brain (eye movements, scalp muscles, line noise, etc.) as well as any near-field effects created at the interface between the electrode and the scalp. 
 
 Use of ICA decomposition to detect and remove (non-brain source) artifacts from EEG and other electrophysiological data has become quite common. It is more slowly becoming common to use ICA decomposition to identify and localize effective cortical sources contributing to the recorded scalp signals that must represent their linear mixtures. EEG signals from the brain itself may be presumed to arise from  near-coherent field activity emerging in densely connected “patches” of cortical tissue. It may be further assumed that the cells contributing directly to the EEG activity are pyramidal cells, as these are arranged roughly perpendicular to the surface of the cortex, their coherent activity thus producing a far-field scalp signal. In the absence of spatially coherent field activities, far field projections from smaller cell populations should cancel each other out when their volume conducted potentials are summed at the scalp electrodes, resulting in quite little far field signal being recorded in the EEG record. 
 
 However, if a cortical patch of sufficient size acts coherently (or even near coherently) its summed activity will be volume conducted to the scalp, producing recordable EEG signal. As such effective source signals from non-coupled cortical areas will tend to be temporally near-independent from each other, ICA decomposition can separate their signals and also identify their distinct (though typically highly overlapping) scalp project patterns, and these in turn can be used to estimate the source location or cortical patch giving rise to the activity identified by ICA decomposition as arising from a single effective source.
 
The spatiotemporal dynamics of local field synchronization in the cortex is an emerging area of neuroscience research. Earlier studies have suggested modeling such field patterns as “phase cones” or “neuronal avalanches.” However, despite the widespread use of ICA decomposition in processing EEG data, few direct comparisons have been published of the relative performance of various ICA and related BSS algorithms.

\subsection{Picard versus other ICA algorithms}
By default, Picard solves the same problem as Infomax, but using a different optimization method. Here we saw that the decompositions produced by Picard do have similar performance in terms of MIR, remnant PMI, and dipolarity, to those produced by Infomax. This aligns with expectations as both algorithms are solving the same optimization problem.

Further, figure \ref{dataset_comp} shows that the ordering of mean MIR produced by the tested algorithms is remarkably consistent across all 13 datasets, although the MIR returned by the decompositions is quite different for the different datasets. This result makes possible the remarkably linear trend shown by the regression (\ref{fig:data}) of the mean MIR values across all datasets for each algorithm versus the percent near-dipolar components. Note that whereas MIR is computed only from the time courses of the returned component activations versus the raw channel data, the 'dipolarity' (percentage of near-dipolar components) is computed only from the component scalp maps learned from the data by algorithms that themselves are given no information about electrode locations, Maxwell equations, etc. Thus, the linearity we again observe here cannot be predicted mathematically, but must arise from the nature of the biological origins of the effective source processes as we have suggested above.

Picard-O outperforms FastICA slightly in terms of MIR, remnant PMI, and dipolarity. However this may be the result of the fact that by default EEGLAB uses a relatively large default FastICA stopping rule step size threshold of $10^{-4}$.

Exploration of the performance of Picard in comparison to its stopping rule threshold here appears to show an ideal stopping rule threshold of $10^{-6}$ as still smaller examined thresholds ($10^{-7}$,$10^{-8}$) do not appear to give higher MIR. The performance of Picard, as measured by MIR, appears to be relatively tolerant of differences in stopping rule thresholds, achieving results similar to Pearson with a much lower stopping rule threshold ($10^{-1}$), and results closer to Infomax with a higher stopping tolerance ($10^{-3}$). This gives an interesting option to users who may wish to trade off higher MIR with quicker computation times, or to test using the result of brief decomposition with Picard as input to computationally more complex but most effective AMICA decomposition, as our results (Fig. 2a) show a clear performance advantage to AMICA relative to the other 19 algorithms tested.

Here Picard and Picard-O both demonstrated somewhat lower run times than Extended Infomax (though not lower than non-extended Infomax), while achieving similar performance in terms of dipolarity and MIR, making Picard and Picard-O both potentially attractive alternatives to Extended Infomax in applications in which computation time is at a premium. In future work, it would be of interest to perform a more in depth comparison of the performance of Picard, Picard-O, Infomax, Extended Infomax, and AMICA in terms of MIR, PMI, and percent near-dipolar components  across stopping rule step size thresholds. Although as a general rule, using a given stopping rule threshold for Picard-O is roughly equivalent to using its square root for FastICA, there seems to be no simple way to compare the stopping tolerances between Infomax and Picard.

\section{Conclusion}
Here a framework for comparing multiple different ICA algorithms introduced by Delorme et al., (\cite{delorme2012independent}) has been again demonstrated. We have also made it openly accessible along with the test data used here on github (https://github.com/sccn/testica). This framework makes use of three metrics to evaluate the relative performance of ICA algorithms on electrophysiological data: Mutual Information Reduction (MIR), Pairwise Mutual Information (PMI), and 'dipolarity'.

Picard is a relatively new ICA algorithm of interest for the somewhat greater speed at which it can solve source separation problems compared to similar algorithms. Here we applied our ICA performance ranking tools to Picard and its orthogonal variant, Picard-O. We confirmed that Picard generally shows relatively comparable performance to Infomax. This is the expected result as Picard solves the same optimization problem as Infomax using a different optimization technique. We further showed that a stopping tolerance of $10^{-6}$ is likely the optimal stopping tolerance to be used with Picard, although nearly equivalent performance may be produced by looser stopping rule thresholds as large as $10^{-3}$.

We hope that making the code and data used here to profile the performance of ICA algorithms public will allow future investigators to be easily able to compare the performance of future ICA algorithms to the performance of existing algorithms as applied to EEG or related data.

\bibliographystyle{IEEEtran}
\bibliography{IEEEabrv,refs}

\begin{thebibliography}{10}
\providecommand{\url}[1]{#1}
\csname url@samestyle\endcsname
\providecommand{\newblock}{\relax}
\providecommand{\bibinfo}[2]{#2}
\providecommand{\BIBentrySTDinterwordspacing}{\spaceskip=0pt\relax}
\providecommand{\BIBentryALTinterwordstretchfactor}{4}
\providecommand{\BIBentryALTinterwordspacing}{\spaceskip=\fontdimen2\font plus
\BIBentryALTinterwordstretchfactor\fontdimen3\font minus
  \fontdimen4\font\relax}
\providecommand{\BIBforeignlanguage}[2]{{%
\expandafter\ifx\csname l@#1\endcsname\relax
\typeout{** WARNING: IEEEtran.bst: No hyphenation pattern has been}%
\typeout{** loaded for the language `#1'. Using the pattern for}%
\typeout{** the default language instead.}%
\else
\language=\csname l@#1\endcsname
\fi
#2}}
\providecommand{\BIBdecl}{\relax}
\BIBdecl

\bibitem{nunez1974brain}
P.~L. Nunez, ``The brain wave equation: a model for the eeg,''
  \emph{Mathematical Biosciences}, vol.~21, no. 3-4, pp. 279--297, 1974.

\bibitem{varela2001brainweb}
F.~Varela, J.-P. Lachaux, E.~Rodriguez, and J.~Martinerie, ``The brainweb:
  phase synchronization and large-scale integration,'' \emph{Nature reviews
  neuroscience}, vol.~2, no.~4, pp. 229--239, 2001.

\bibitem{stepanyants2009fractions}
A.~Stepanyants, L.~M. Martinez, A.~S. Ferecsk{\'o}, and Z.~F. Kisv{\'a}rday,
  ``The fractions of short-and long-range connections in the visual cortex,''
  \emph{Proceedings of the National Academy of Sciences}, vol. 106, no.~9, pp.
  3555--3560, 2009.

\bibitem{stettler2002lateral}
D.~D. Stettler, A.~Das, J.~Bennett, and C.~D. Gilbert, ``Lateral connectivity
  and contextual interactions in macaque primary visual cortex,''
  \emph{Neuron}, vol.~36, no.~4, pp. 739--750, 2002.

\bibitem{sarnthein2005thalamocortical}
J.~Sarnthein, A.~Morel, A.~Von~Stein, and D.~Jeanmonod, ``Thalamocortical theta
  coherence in neurological patients at rest and during a working memory
  task,'' \emph{International journal of psychophysiology}, vol.~57, no.~2, pp.
  87--96, 2005.

\bibitem{dehghani2010magnetoencephalography}
N.~Dehghani, S.~S. Cash, A.~O. Rossetti, C.~C. Chen, and E.~Halgren,
  ``Magnetoencephalography demonstrates multiple asynchronous generators during
  human sleep spindles,'' \emph{Journal of neurophysiology}, vol. 104, no.~1,
  pp. 179--188, 2010.

\bibitem{delorme2012independent}
A.~Delorme, J.~Palmer, J.~Onton, R.~Oostenveld, and S.~Makeig, ``Independent
  eeg sources are dipolar,'' \emph{PloS one}, vol.~7, no.~2, p. e30135, 2012.

\bibitem{ablin2018faster}
P.~Ablin, J.-F. Cardoso, and A.~Gramfort, ``Faster independent component
  analysis by preconditioning with hessian approximations,'' \emph{IEEE
  Transactions on Signal Processing}, vol.~66, no.~15, pp. 4040--4049, 2018.

\bibitem{chaisaen2020decoding}
R.~Chaisaen, P.~Autthasan, N.~Mingchinda, P.~Leelaarporn, N.~Kunaseth,
  S.~Tammajarung, P.~Manoonpong, S.~C. Mukhopadhyay, and T.~Wilaiprasitporn,
  ``Decoding eeg rhythms during action observation, motor imagery, and
  execution for standing and sitting,'' \emph{IEEE sensors journal}, vol.~20,
  no.~22, pp. 13\,776--13\,786, 2020.

\bibitem{sangnark2021revealing}
S.~Sangnark, P.~Autthasan, P.~Ponglertnapakorn, P.~Chalekarn,
  T.~Sudhawiyangkul, M.~Trakulruangroj, S.~Songsermsawad, R.~Assabumrungrat,
  S.~Amplod, K.~Ounjai \emph{et~al.}, ``Revealing preference in popular music
  through familiarity and brain response,'' \emph{IEEE Sensors Journal},
  vol.~21, no.~13, pp. 14\,931--14\,940, 2021.

\bibitem{kudo2022magnetoencephalography}
K.~Kudo, H.~Morise, K.~G. Ranasinghe, D.~Mizuiri, A.~S. Bhutada, J.~Chen,
  A.~Findlay, H.~E. Kirsch, and S.~S. Nagarajan, ``Magnetoencephalography
  imaging reveals abnormal information flow in temporal lobe epilepsy,''
  \emph{Brain Connectivity}, vol.~12, no.~4, pp. 362--373, 2022.

\bibitem{liu1989limited}
D.~C. Liu and J.~Nocedal, ``On the limited memory bfgs method for large scale
  optimization,'' \emph{Mathematical programming}, vol.~45, no.~1, pp.
  503--528, 1989.

\bibitem{ablin2018ortho}
P.~Ablin, J.-F. Cardoso, and A.~Gramfort, ``Faster ica under orthogonal
  constraint,'' in \emph{2018 IEEE International Conference on Acoustics,
  Speech and Signal Processing (ICASSP)}.\hskip 1em plus 0.5em minus
  0.4em\relax IEEE, 2018, pp. 4464--4468.

\bibitem{onton2005frontal}
J.~Onton, A.~Delorme, and S.~Makeig, ``Frontal midline eeg dynamics during
  working memory,'' \emph{Neuroimage}, vol.~27, no.~2, pp. 341--356, 2005.

\bibitem{delorme2004eeglab}
A.~Delorme and S.~Makeig, ``Eeglab: an open source toolbox for analysis of
  single-trial eeg dynamics including independent component analysis,''
  \emph{Journal of neuroscience methods}, vol. 134, no.~1, pp. 9--21, 2004.

\end{thebibliography}

\end{document}